\documentclass[12pt]{iopart}

\usepackage{epsfig}
\usepackage{iopams}

\begin{document}

\title[Cooperative effect of load and disorder in thermally activated rupture]
{Cooperative effect of load and disorder in thermally activated
rupture of a $2d$ random fuse network}
\author{A. Guarino$^1$, L. Vanel$^2$, R. Scorretti$^3$ and S. Ciliberto$^2$}
\address{$^1$ Universit\'e de la Polyn\'esie Fran\c{c}aise, BP 6570
Fa\`{a}a Aeroport, Tahiti, French Polynesia}
\address{$^2$
Laboratoire de physique, CNRS UMR 5672, Ecole Normale Sup\'erieure
de Lyon, 46 all\'ee d'Italie, 69364 Lyon Cedex 07, France}
\address{$^3$ CEGELY - UMR 5005, Universit\'e Lyon, 1 43, bld du 11 Nov.
1918, 69622 Villeurbanne Cedex, France} \ead{Loic.Vanel@ens-lyon.fr}

%

\begin{abstract}
A random fuse network, or equivalently a $2d$ spring network with
quenched disorder, is submitted to a constant load and thermal
noise, and studied by numerical simulations. Rupture is thermally
activated and the lifetime follows an Arrhenius law where the energy
barrier is reduced by disorder. Due to the non-homogenous
distribution of forces from stress concentration at microcracks'
tips, spatial correlations between rupture events appear, but they
do not affect the energy barrier's dependence on disorder, they
affect only the coupling between disorder and the applied load.
\end{abstract}
\bigskip

The effect of quenched disorder on dynamics is a recurring problem
in many physical systems with elastic interactions. The motion of
vortex lines in supraconductors, charge-density waves in Bragg
glasses, magnetic domains walls, or contact lines in wetting show a
competition between elastic interactions and pinning by disorder
\cite{Kardar,Giamarchi}. While many studies have focused on systems
driven above a critical depininng threshold, an important issue
remains to understand the sub-critical regime, when thermally
activated creep motion occurs \cite{Rosso,Brazovski}.

Rupture in disordered brittle solids falls in the same class of
problems \cite{Fisher}. Elastic interactions tend to make a crack
propagate in a straight direction while disorder creates roughness
\cite{Bouchaud} or causes spatially diffuse damage
\cite{HerrmannRoux,Garcimartin}. In the sub-critical rupture regime,
a very important quantity for safety reasons is the lifetime, i.e.
the mean time for a sample to break under a prescribed load. The
lifetime follows an Arrehnius law \cite{Brenner,Zhurkov}, but
thermal noise is generally too small compared to recent theoretical
estimates of the energy barrier \cite{Golubovic,Pomeau,Buchel} to
explain experimental observations in heterogenous materials
\cite{Guarino,Santucci,Rabinovitch04}. For athermal systems,
disorder actually reduces the energy barrier and can be seen as an
effective temperature \cite{Arndt}. In order to clarify the role of
disorder in thermal systems, one-dimensional Thermal and Disordered
Fiber Bundles Models ($1d$-TDFBM) have been introduced to model the
thermally activated rupture of an heterogenous material submitted to
a constant external load
\cite{Roux00,Scorretti01,Ciliberto01,Politi02,Sornette05}. The TDFBM
considers an elastic system in equilibrium at \textit{constant
temperature} where statistical force fluctuations occur in time due
to thermal noise. This is very different from previous thermal
random fuse models \cite{Sornette92} where rupture results from an
increase in fuse temperature due to dissipation through a
generalized Joule effect until reaching a critical melting
temperature.  In a TDFBM, elastic energy is the equivalent of
dissipation in the thermal random fuse model but does not cause
rupture when the system is at mechanical equilibrium; rupture is
caused by elastic force fluctuations analogous to Nyquist noise.
%

One problem with the $1d$-TDFBM investigated up to now is that the
load is shared equally among all the unbroken fibers. This is not a
realistic load-sharing rule for experimental geometries where
elasticity cause stress concentration at microcracks' tips and lead
to a non-uniform redistribution of stress. In this letter, we show
that spatial correlations between rupture events in $2d$ do not
affect the dependence of the energy barrier on disorder but only the
coupling between disorder and the applied load.

First, we discuss briefly results obtained by several authors
\cite{Roux00,Scorretti01,Ciliberto01,Politi02, Sornette05} on the
$1d$-TDFBM. The system considered is made of a set of $N$ parallel
fibers, each carrying an initial force $f_0$ and behaving as a
linear elastic spring with unity stiffness. Each fiber $j$ can carry
a maximum force $f^{(j)}_c$ before it breaks. Quenched disorder is
introduced in the system by distributing thresholds $f^{(j)}_c$
according to a gaussian distribution of mean $<f^{(j)}_c>=1$ and
variance $T_d$; for each fiber, the value $f^{(j)}_c$ is a
time-independent constant. Contrary to the case of non-thermal
$1d$-DFBM where the system evolves due to a progressive increase in
total current, we consider that the total force applied to the
$1d$-TDFBM is kept constant. Dynamics is introduced in the system by
introducing fluctuations in spring forces due to thermal noise. We
write $f_j$ the average force on fiber $j$. The fluctuations in
force $\delta f_j$ that occur in time on fiber $j$ are assumed to
follow a gaussian probability distribution with $0$ mean value and
variance $T$, where $T$ represents the thermodynamical temperature
in unit of square force. When the total force on a fiber $f_j+\delta
f_j$ is larger than the threshold $f^{(j)}_c$, the fiber breaks. The
remaining fibers share equally the total force: this is a so-called
democratic model. The bundle will break completely as soon as the
average force on each fiber exceeds the breaking threshold. Roux has
shown that the mean time to break the first fiber follows an
Arrhenius law where disorder acts as an additive temperature
\cite{Roux00}:
\begin{equation}
\tau \sim \exp\left(\frac{(1-f_0)^2}{2(T+T_d)}\right) \label{eqU1}
\end{equation}
with $f_0$ the initial force carried by each fiber of the bundle. In
the general case where many fibers break before total rupture of the
bundle, the lifetime obeys an Arrhenius law with a different general
form :
\begin{equation}
\tau \sim \exp\left(\frac{U(f_0,T_d)}{T}\right) \label{eqTau}
\end{equation}
An approximate expression of $U$ for low disorder is
\cite{Scorretti01,Ciliberto01}:
\begin{equation}
U=\frac{1}{2}\left(1-f_0-\beta_1\sqrt{T_d}\right)^2
\label{eqUapprox}
\end{equation}
with $\beta_1=\sqrt{\pi/2}$. For higher disorder, Politi et al
\cite{Politi02} have shown numerically that $U$ is determined with a
very good approximation by the minimum value of the rate at which
fibers break. More precisely, if $n$ is the number of broken fibers
and $N$ the total number of fibers then $\Phi= n/N$ is the fraction
of broken fibers and $\dot{\Phi}$ its time derivative. The value
$\Phi^*$ for which $\dot{\Phi}$ is minimum obeys an implicit
equation \cite{Politi02, Sornette05}:
\[\exp\left[\rm{irfc}^2(2\Phi^*)\right](1-\Phi^*)^2=\alpha\]
where $\rm{irfc}$ is the inverse function of the complementary error
function and $\alpha=f_0/\sqrt{2\pi T_d}$. Then, $U$ can be
approximated as \cite{note}:
\begin{equation}
\fl U = \frac{1}{2}\left(1-f_0-\beta_2 \sqrt{T_d}\right)^2
\qquad\rm{with}\qquad \beta_2(\alpha)=
\alpha\sqrt{2\pi}\frac{\Phi^*}{1-\Phi^*}+\sqrt{2}\,\rm{irfc}(2\Phi^*)
\label{eqUexact}
\end{equation}
Note that the coupling coefficient $\beta_2$ depends on both
disorder $T_d$ and applied load $f_0$. Eq.(\ref{eqUexact}) predicts
that the variation of $U$ with $T_d$ for a fixed $f_0$ is non
monotonous \cite{Sornette05}. However, when $\alpha > 1/4$ (this
condition corresponds to $\Phi^* < 1/2$ which would be expected in
practice for most materials), $U$ decreases when $T_d$ is increased
with $f_0$ constant. In that case, the function $\beta_2$ has a
lower value $\beta_2(1/4)=\beta_1/2$ and an asymptotic behavior for
large $\alpha$ ($\Phi^* \rightarrow 0$),
$\beta_2(\alpha)\simeq\sqrt{2 \ln \alpha}+1/\sqrt{2 \ln \alpha}$
\cite{Politi02}.

To study the effect of a non-uniform force redistribution on the
rupture dynamics of the TDFBM, one could keep a $1d$ geometry and
introduce a finite range of interaction between fibers
\cite{Herrmann}. Instead, we work directly in a $2d$ geometry more
closely related to a real experiment. The above described 1d-TDFBM
is equivalent to a system of parallel fuses where forces are
transformed in currents and displacements in electric potentials. A
$2d$ square fuse network is then equivalent to a square lattice of
springs in antiplane deformation. Specifically, each node of the $N
\times N$ nodes square lattice can move along an axis perpendicular
to the plane of springs at rest. A constant force $F$ is applied at
two opposite sides of the lattice in antiplane configuration. In the
initial equilibrium configuration of the lattice, the springs
submitted to a load $f_0=F/N$ are called "parallel" springs while
the unloaded springs (zero force) are called "orthogonal" springs.

Like in $1d$, the fluctuations in force $\delta f_j$ on spring $j$
follow a gaussian probability distribution with $0$ mean value and
variance $T$, and the rupture thresholds $f_c^{(j)}$ follow a
gaussian distribution of mean $<f^{(j)}_c>=1$ and variance $T_d$.
The time scale in the simulation is the (constant) time between two
configurations of force fluctuations in the system. The whole
network is a square with sides 100 springs wide; thus, the lattice
contains about $2.\,10^4$ springs. Whenever we choose rupture
thresholds from the gaussian distribution, there is a non-zero
probability to obtain a negative threshold : $P_{<0}
=\frac{1}{2}\rm{erfc}\left(<f_c>/\sqrt{2 T_d}\right)$. For a system
with $2\,10^4$ springs, we can safely consider that no spring will
spontaneously break when there is no load ($f_0=0$) and at zero
temperature if $P_{<0} < 10^{-5}$, thus $T_d < 0.055$ when
$<f_c>=1$. In practice, we will work with $T_d<0.05$.

First, let us consider a lattice with no disorder in the rupture
thresholds and all the springs initially intact. When thermal noise
is on, some springs will start to break. Due to stress concentration
effects, as soon as one of the parallel springs is broken, the
neighboring springs are submitted to a larger force. If that new
force exceeds the rupture threshold then the neighboring spring will
also break, and the force on the next neighbor will be even higher.
This process will result in the rupture of the whole network in an
avalanche started from a single rupture event. Numerically, this
will occur in our lattice as soon as $f_0 > f_c^1$, where $f_c^1
(\simeq 0.785)$ is the critical threshold of the homogenous lattice
at $T=0$ when one parallel spring is broken. In that case, the
rupture time will be directly related to the probability of breaking
a single spring in the network and, if the lattice is disordered, we
will recover essentially the result given by eq.~(\ref{eqU1})
\cite{Roux00}. In the rest of the paper, we will be interested only
in the case where several springs break before the final avalanche
occurs, i.e. when $f_0 < f_c^1$.

\begin{figure}[h]
\centerline{\includegraphics[width=8cm]{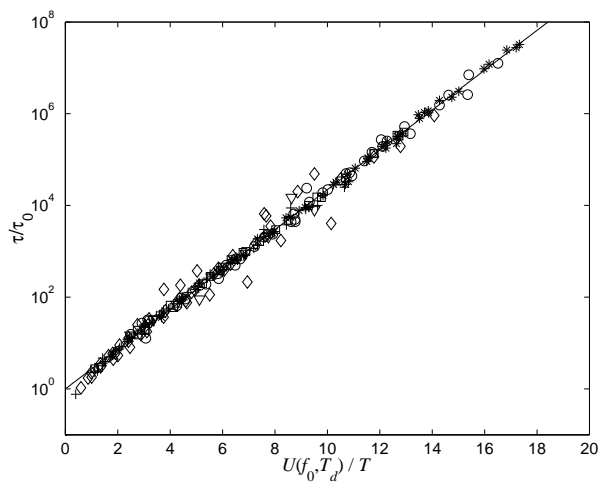}} \caption{Scaled
rupture time versus Arrhenius factor $U/T$. Various symbols
correspond to different values of disorder and applied force. The
solid line corresponds to $\log y = x$.} \label{tau}
\end{figure}

For a fixed value of the applied force $f_0$ and a fixed value of
the disorder $T_d$, we measure as a function of temperature $T$ the
lifetime, i.e. the rupture time averaged over as many as $100$
numerical experiments. We find that lifetime follows an Arrhenius
law and obtain numerically $U(f_0,T_d)$ as defined by
eq~(\ref{eqTau}). We see on figure~\ref{tau} the collapse of the
data for a range of values $0.04 < f_0 < 0.77 $ and
$10^{-4}<T_d<0.05$. Some of the data points are more scattered
around the expected behavior (solid line) than others because the
ratio of the standard deviation over the mean of the rupture time
increases when $T$ becomes small. This property, already mentioned
in \cite{Roux00}, makes numerical convergence of the mean difficult
in some cases.

\begin{figure}[htbp]
\centerline{\includegraphics[width=8cm]{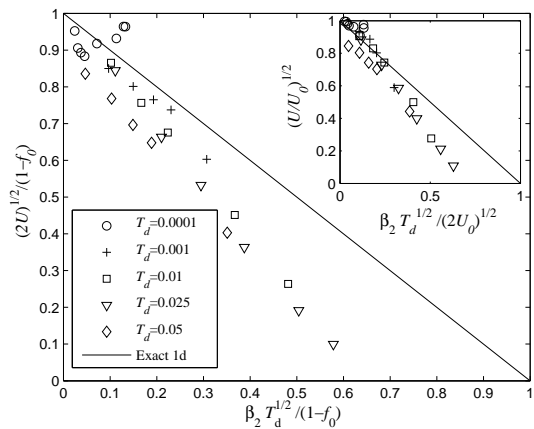}}
\caption{Comparison between the numerical value of $U$ in $2d$ and
eq.~(\ref{eqUexact}) in $1d$ (solid line). Inset : same scaling, but
using the numerical value $U_0=U(f_0,0)$ for the $2d$ lattice.}
\label{Uexact}
\end{figure}

\begin{figure}[htbp]
\centerline{\includegraphics[width=7.6cm]{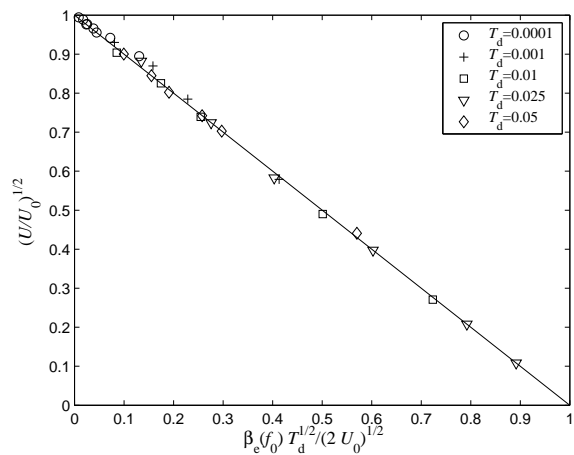}}
\caption{Collapse of all the numerical data using an effective value
$\beta_e$ which is only a function of $f_0$ but not of $T_d$,
contrary to the case of $\beta_2$ which is a function of $T_d$.}
\label{Uapp}
\end{figure}

\begin{figure}[htbp]
\centerline{\includegraphics[width=8cm]{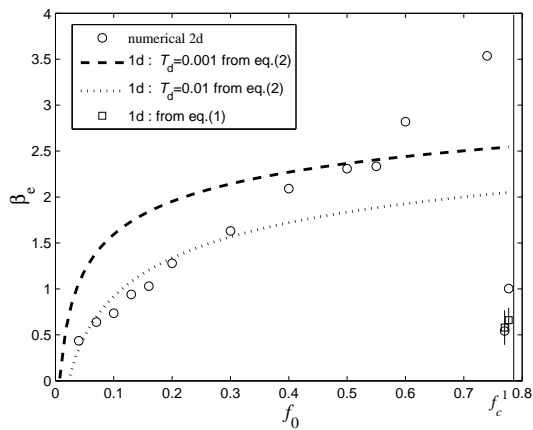}}
\caption{Numerical values of $\beta_e$ as a function of $f_0$
(circles). The dashed and dotted lines correspond to $\beta_2$ as
predicted by eq.~(\ref{eqUexact}) when $T_d=0.001$ and $T_d=0.01$.
Close to $f_c^1$, the values found for $\beta_e$ are close to values
predicted by eq.~(\ref{eqU1}).} \label{beta}
\end{figure}

To compare the barrier $U(f_0,T_d)$ for the $2d$ geometry with the
one of the $1d$-TDFBM, we plot on figure~\ref{Uexact}
$\sqrt{2U}/(1-f_0)$ as a function of $\beta_2 \sqrt{T_d}/(1-f_0)$.
We clearly see that the functional form for $\beta_2$ is not
correct, but also that the data for various values of disorder do
not rescale very well. The first immediate reason for the
discrepancy is that for $T_d=0$, the energy barrier of the $2d$
network $U_0=U(f_0,0)$ (determined numerically) is not $(1-f_0)^2/2$
as in $1d$. This is due to the preferential redistribution on the
nearest neighbors of the force carried by a fiber before rupture.
Taking into account the effective energy barrier in $2d$ does not
improve the comparison with the 1d-TDFBM. After replacing $1-f_0$ by
$\sqrt{2 U_0}$ in eq.~(\ref{eqUexact}), we see in the inset of
figure~\ref{Uexact} that it is not enough to get a good collapse of
all the data on the theoretical prediction (solid line).

We find that the barrier decreases with disorder following the
linear curve : $\sqrt{U(f_0,T_d)}=\sqrt{U_0}-b\sqrt{T_d}$ for
$10^{-4}<T_d<0.05$ and a fixed value of $f_0$. Thus, it turns out
that eq.~(\ref{eqUapprox}) is a much better functional form than
eq.~(\ref{eqUexact}), even though it is an approximate expression in
$1d$. Looking at eq.~(\ref{eqUapprox}) or eq.~(\ref{eqUexact}), we
can make an analogy with the $1d$ case and say that the second
coefficient $b$ corresponds to an effective value $\beta_e/\sqrt{2}$
which is now only a function of $f_0$. On figure~\ref{Uapp}, we see
the collapse of all the data when we plot $\sqrt{U/U_0}$ as a
function of $\beta_e(f_0) \sqrt{T_d}/\sqrt{2 U_0}$.

The coupling coefficient $\beta_e$ increases almost linearly with
$f_0$ up to values close to $f_c^1$ (figure~\ref{beta}). However,
when $f_0$ gets very close to $f_c^1$, there is an abrupt decrease
in the value of $\beta_e$. This is related to the fact that rupture
is now controlled by a single event as in eq.~(\ref{eqU1}). Indeed,
although eq.~(\ref{eqU1}) does not follow the general scaling
property of eq.(\ref{eqTau}), we can estimate a value $\beta_e$ for
each temperature value used in the simulation. The average value
found for $\beta_e$ from eq.~(\ref{eqU1}) (square symbols in
figure~\ref{beta} with an error bar corresponding to variations with
temperature) is a reasonable estimate of the numerical value.

The functional behavior of $\beta_e$ is very different from that of
the $1d$ model (eq.~(\ref{eqUexact})) where $\beta_2$ depends on
both $f_0$ and $T_d$. As an example, we plot on figure~\ref{beta}
$\beta_2(f_0)$ for fixed values of $T_d$. Not only the functional
dependence is clearly different from the numerical estimate
$\beta_e(f_0)$ but also $\beta_2$ decreases with $T_d$ at fixed
$f_0$. In that sense, the load and the disorder do not act
cooperatively in $1d$.

\begin{figure}[htbp]
\centerline{\includegraphics[width=8cm]{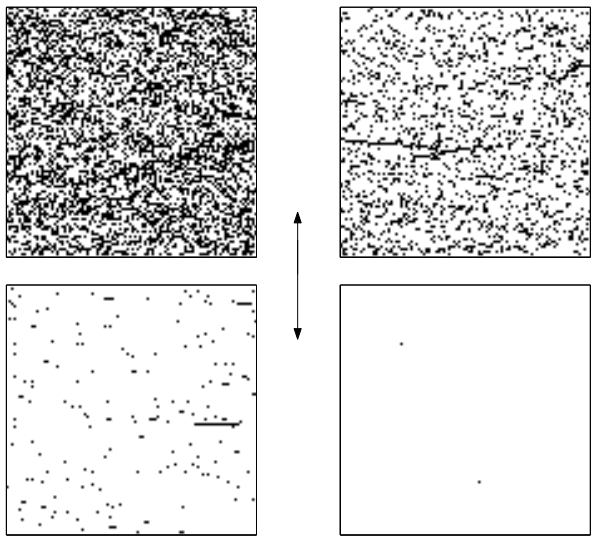}} \caption{Image of
the broken fibers (black dots) just before the final rupture for
$T_d=0.01$ and $f_0=0.05$ (top left), $f_0=0.1$ (top right),
$f_0=0.3$ (bottom left); $T_d=0.004$ and $f_0=0.75$ (bottom right).
The arrow shows the loading direction.} \label{Image}
\end{figure}

The key point in $2d$ is that the spatial correlations between
rupture events depend on the strength of stress intensification.
This is illustrated on figure~\ref{Image} a) to c) which shows the
broken fibers just before the final avalanche for different values
of $f_0$ and a fixed value of $T_d$. For very small loads, damage is
scattered everywhere in the sample. At higher loads, damage becomes
less scattered and growth of straight cracks occurs. Finally,
figure~\ref{Image}d) shows that for a load close to the critical
threshold, only very few events occur. A similar transition was
observed for zero disorder or annealed disorder models with power
law rate of rupture \cite{Hansen,Curtin,Newman}. In contrast to
these models where the transition occurs by changing the exponent of
the power law, we observe a transition resulting from the
competition between stress intensification and quenched disorder.

To understand this transition in our model, let us consider the
increase in force due to stress intensification when a spring
breaks. If the increase is small compared to $\sqrt{T_d}$, there
will be very little spatial correlation between rupture events
occurring preferentially at the weakest springs. For a given
disorder, there is always a force $f_0$ small enough to observe this
rupture regime similar to the $1d$-TDFBM case. On the contrary, if
the increase due to stress intensification is large compared to
$\sqrt{T_d}$, it is easier to break a spring next to an already
broken spring, and the rupture will proceed mainly by growth of
multiple cracks. In spite of very different regimes of spatial
correlation between rupture events, we have the remarkable result
that the energy barrier dependence on $T_d$ is unchanged. Spatial
correlations only affects the coupling coefficient $\beta_e$,
increasing quasi-linearly with $f_0$ and independent of $T_d$.

The multiplicative amplification of disorder due to $\beta_e$ is a
mechanism that will create a load-dependent reduction of the energy
barrier in thermally activated rupture. It will have an effect on
the order of magnitude and load-dependence of the rupture time which
could help understanding experiments in heterogeneous materials
\cite{Guarino}.

 In conclusion, we have studied thermally activated rupture of a
$2d$ elastic spring network submitted to a constant load and thermal
noise. We find that spatial correlations between rupture events are
controlled by a competition between quenched disorder and force
inhomogeneities due to stress concentration. For low spatial
correlations, the energy barrier scales naturally like in the $1d$
model. Remarkably, the appearance of spatial correlations does not
affect the functional dependence of the energy barrier on disorder,
but only the coupling coefficient $\beta_e$ which is independent of
disorder and increases quasi-linearly with the applied load $f_0$.
This is an important result showing that the applied load contribute
to amplify in a cooperative way the effect of disorder on the
lifetime. The observed cooperative effects of load and disorder in
$2d$ subcritical rupture could be relevant to the creep regime of
other physical systems with elastic interactions
\cite{Kardar,Giamarchi} and also to crackling noise \cite{Dahmen}.


\begin{thebibliography}{unsrt}
\bibitem{note} This is the correct expression of $U$. It has been published with several typographic mistakes as eq. (35)
in \cite{Sornette05}.
\end{thebibliography}
\end{document}